\documentclass[12pt]{article}

\usepackage{graphicx}
\usepackage{amsmath,amssymb,amsfonts}
\usepackage{hyperref}
\usepackage{lmodern}
\usepackage{amssymb}
\usepackage{booktabs}

\renewcommand{\theequation}
{\arabic{section}.\arabic{equation}}

\makeatletter
\def\eqnarray{ \stepcounter{equation} \let\@currentlabel=\theequation
 \global\@eqnswtrue
 \global\@eqcnt\z@
 \tabskip\@centering
 \let\\=\@eqncr
 $$\halign to \displaywidth\bgroup\@eqnsel\hskip\@centering
 $\displaystyle\tabskip\z@{##}$&\global\@eqcnt\@ne
 \hfil$\displaystyle{{}##{}}$\hfil
 &\global\@eqcnt\tw@$\displaystyle\tabskip\z@{##}$\hfil
 \tabskip\@centering&\llap{##}\tabskip\z@\cr}
\makeatother

\makeatletter
\def\@arrayacol{\edef\@preamble{\@preamble \hskip .5\arraycolsep}}
\def\array{\let\@acol\@arrayacol \let\@classz\@arrayclassz
\let\@classiv\@arrayclassiv \let\\\@arraycr\def\@halignto{}\@tabarray}
\makeatother
 
\newcommand{\be}{\begin{equation}}
\newcommand{\ee}{\end{equation}}

\newcommand{\beqa}{\begin{eqnarray}}
\newcommand{\eeqa}{\end{eqnarray}}
\newcommand{\nn}{\nonumber}

\def\CD {{\cal D}}

\def\CO {{\cal O}}

\begin{document}

\setlength{\baselineskip}{7mm}

\hfill  NRCPS-HE/2020-68

\vspace{1cm}
\begin{center}
{\Large ~\\{\it  Gravity with Perimeter  Action \\

\vspace{0.3cm}
and \\
\vspace{0.3cm}
Gravitational Singularities 

}

}

\vspace{2cm}

{\sl   George Savvidy

\bigskip
  {\sl Institute of Nuclear and Particle Physics\\
  Demokritos National Research Center\\
 Ag. Paraskevi,  Athens, Greece}

\bigskip

} 
\end{center}
\vspace{1cm}

\centerline{{\bf Abstract}}

\vspace{0.5cm}

\noindent
We consider the perturbation of the Schwarzschild solution by the perimeter action. The asymptotic behaviour of the solution at infinity and at the horizon are calculated and analysed in the first approximation. In the regions far from the matter sources the perturbations are characterised by the ratio of the Plank length to the Schwarzschild radius and are infinitesimally small.  At short distances the perturbation is large and there appears a space-time region of the Schwarzschild radius scale  that is unreachable by test particles. These regions are located there where the standard theory of gravity has singularities.

\newpage

\pagestyle{plain}

\section {\it Perimeter  Action }

Unification of gravity with other fundamental forces within the superstring theory stimulated the interest to the quantum gravity and physics at Planck scale 
\cite{sacharov,Buchdahl,Starobinsky:1980te,Adler:1982ri,Gasperini:1992em}. In particular, string theory predicts a modification of the gravitational action 
at Planck scale with additional high-derivative terms. This allows to ask fundamental questions concerning physics at Planck scale referring to these effective actions  and, in particular, one can try to understand how they influence the gravitational singularities \cite{Oppenheimer:1939ue,Finkelstein:1958zz,Penrose:1964wq,Christodoulou,Hawking1,Hawking2,Hawking3,Hawking4,HawkingPenrose,Robertson,Raychaudhure,Komar,Brandenberger:1988aj,Alvarez:1984ee,Baierlein:1962zz,Brandenberger:1993ef,Deser:1998rj} and the black hole physics \cite{Hooft:2015rdz,Gaddam:2020rxb,Dray:1984ha,Hooft:2017bjm,Betzios:2020xuj}. The classical and quantum gravity  theories with generic higher-curvature 
terms  were considered in \cite{Stelle:1976gc,Stelle:1977ry,Deser:2002jk} and recently in \cite{Kehagias:2015ata,Alvarez-Gaume:2015rwa}. The theories with limited-curvature hypothesis were considered in \cite{Markov:1982ed,anini,Ferraro:2006jd}.
 
It is appealing to extend  this approach to different modifications of general relativity that follow from the string theory and also to develop an alternative approach based on new geometrical principles \cite{Savvidy:1995mr,Ambjorn:1996kk,Savvidy:1997qf,Savvidy:2017srm,Savvidy:2018xdf,Ambjorn:1985az,Ambartsumian:1992pz, Ambjorn:1997ub,Savvidy:2015ina}. Here the idea is to extend the Feynman  path integral to an integral over the space-time manifolds in a way that makes  the quantum-mechanical amplitudes proportional to the "linear size" or "perimeter"  of the  four-dimensional  universe \cite{Savvidy:2015ina}.  That will suppress the growth of the lower-dimensional spikes out of a 4-D manifold \cite{Savvidy:1995mr,Ambjorn:1996kk,Savvidy:1997qf,Savvidy:2017srm,Savvidy:2018xdf}.

\begin{figure}
\begin{center}
\includegraphics[width=6cm]{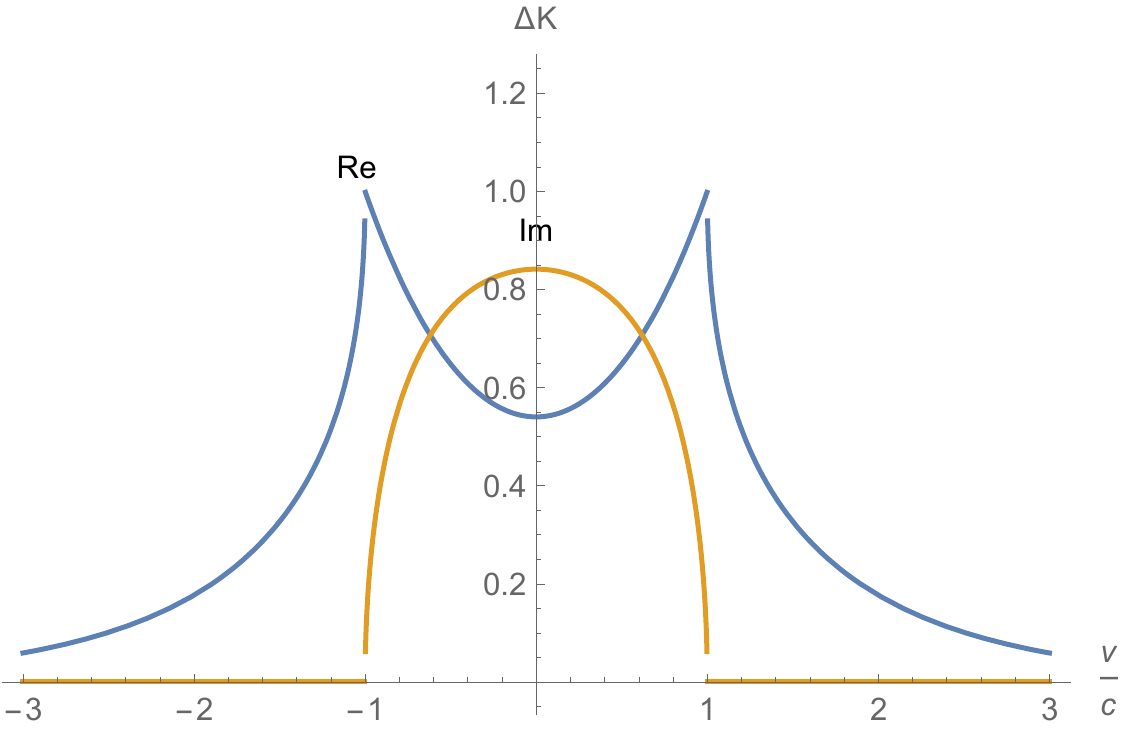}~~~~~~~~~~
\caption{
The graphic of the rial and imaginary parts of the amplitude $ \triangle K =   e^{ {i \over \hbar}  m c^2 \sqrt{1- {v^2 \over c^2}}\triangle t} $.    
}
\label{fig1}
\end{center}
\end{figure}

The suggested  "perimeter" action can be considered as a "square root" of the Regge area action in discretised gravity \cite{Regge:1961px,wheeler}.  In  the Regge action  the {\it area} $\sigma_{ijk}$ of the  
triangle $<ijk>$  of the four-dimensional simplex is  multiplied by the corresponding deficit angle $\omega^{(2)}_{ijk}$ and a summation is over all triangles:
\begin{equation}
S_A= \sum_{<ijk>} \sigma_{ijk} \cdot \omega^{(2)}_{ijk} .
\label{area}
\end{equation}
The formula represents the discretised version of the continuous area action in gravity: 
\be\label{standgravity}
S_A =-{  c^3 \over 16 \pi G } \int R \sqrt{-g} d^4x.
\ee
In the alternative  "perimeter" action \cite{Savvidy:1995mr,Ambjorn:1996kk,Savvidy:1997qf,Savvidy:2017srm,Savvidy:2018xdf} now the perimeter  $\lambda_{ijk}$  of the triangle $<ijk>$  is multiplied by the corresponding deficit angle:
\begin{equation}
S_P = \sum_{<ijk>}\lambda_{ijk} \cdot \omega^{(2)}_{ijk} .
\label{lamb1}
\end{equation}
The Regge action measures  the "area" of the universe, the perimeter action measures the "linear size" of the universe and requires the introduction of the fundamental length scale.
It is unknown to the author how to derive a continuous limit  of the perimeter  action (\ref{lamb1}) in a unique way. In these circumstances one can try to construct a possible perimeter  action for a smooth manifold of a space-time universe by using the available geometrical invariants.  Any expression which is quadratic in the curvature tensor and includes two derivatives can be a candidate for the perimeter action.  These invariants  have the dimension of $1/cm^6$:
\beqa\label{lineargravity2}
 I_1= -{1 \over 80\pi }R_{\mu\nu\lambda\rho;\sigma} R^{\mu\nu\lambda\rho;\sigma} ,~~~~~~~~~~I_2=+  {1 \over 16 \pi }R_{\mu\nu\lambda\rho}  \Box  R^{\mu\nu\lambda\rho}~,  
\eeqa
and we will consider a linear combination of the above invariants\footnote{The general form of the action is presented in the Appendix (\ref{generalform}).}:
 \beqa\label{lineargravity}
 && S_P=    M c 
  \int   \sqrt{I_1 +(1-\epsilon) I_2 }~\sqrt{-g }  d^4x.
\eeqa
The dimension of the integrant invariant is ${1/ cm^3} $, and  the above four-dimensional space-time integral has the dimensionality of $cm$. We have to introduce a mass parameter $M$ to get a correct dimensionality $(g~ cm^2/sec)$ of the action (\ref{lineargravity}). The mass parameter can be expressed in terms of the Planck mass $M_P $ leading to the appearance of the Planck constant $\hbar$ in the action.  We also  introduced a dimensionless coupling constant  $\gamma$ expressing the mass parameter in terms of Planck mass units, 
\be\label{Planckmass}
M  = \gamma ~M_P ,~~~~~~M_P=\sqrt{\hbar c / 16 \pi G},
\ee
thus the perimeter action is: 
\beqa\label{lineargravity1}
 && S_P=    \gamma   \sqrt{\hbar} \ \sqrt{  c^3  \over  16 \pi G}   \int  \sqrt{ I_1+ (1-\epsilon) I_2
}~\sqrt{-g }  d^4x.
\eeqa
The  action (\ref{lineargravity1}) fulfils our basic physical requirement on the action that it should have the dimension of length and should be similar to the action of the relativistic particle \cite{Savvidy:1995mr,Ambjorn:1996kk,Savvidy:1997qf,Savvidy:2017srm,Savvidy:2018xdf}:
\be\label{relatparticle}
S=-m c \int  d s= -m c^2 \int \sqrt{1-{\vec{v}^2 \over c^2 }}~ dt.
\ee
Both expressions contain the geometrical invariants that are not in general positive-definite under the square root operation. In the relativistic particle case (\ref{relatparticle}) the expression under the root becomes negative for a particle moving with a velocity that exceeds the velocity of light. In that case the action develops an imaginary part, and the quantum-mechanical superposition of the amplitudes prevents a particle from exceeding the velocity of light \cite{Pauli:1941zz,Feynman:1949hz,Feynman:1949zx} as it is demonstrated in Fig.\ref{fig1}.  A similar mechanism  was implemented in the Born-Infeld modification of electrodynamics with the aim to prevent the appearance of infinitely large electric fields \cite{borninfeld,Markov:1982ed,anini,Ferraro:2006jd}.
One can expect that in the case of the perimeter action (\ref{lineargravity}) there may appear space-time regions that are unreachable by the test particles as far as in that regions the action develops an imaginary part and the amplitude is exponentially small as it can be seen in Fig.\ref{fig2}.  If these  space-time regions happen  to appear and if space-time regions include singularities, then one can expect  that the gravitational singularities are naturally excluded from the theory  due to the fundamental principles of the quantum mechanics. The question of consistency  of the new action principle, if it is  the right one, can only be decided by its physical consequences. 

In the next sections we will consider the  black hole (BH) singularities and the physical effects that are induced  by the inclusion of the perimeter action (\ref{lineargravity}). As
we will see, the expression under the root in (\ref{lineargravityS}) becomes negative in the region that is smaller than the Schwarzschild radius $r_g$ and includes the singularities. For the observer that is far away from the BH horizon the perimeter action induces a tiny advance precession of the perihelion but has a profound influence on the physics near the horizon. We are confronted here with space-time regions that are unreachable by the test particles and  the expectation value of any observable $\langle  \CO \rangle$ in that region will be exponentially suppressed (\ref{lineargravityS}), (\ref{fullaction}) and (\ref{amplitude}). If one accepts this concept, then it seems plausible that the gravitational singularities are  excluded from the modified theory. In this letter we have only taken the first steps to describe the phenomena  that are caused by the additional perimeter term in the gravitational action. 

\section{\it Perimeter Perturbation of Schwarzschild  Solution }

The modified action that we will consider is a sum
\beqa\label{actionlineargravity23}
 && S =-{  c^3 \over 16 \pi G  } \int R \sqrt{-g} d^4x 
+ \gamma   \sqrt{\hbar} \ \sqrt{ c^3  \over  16 \pi G}   \int   \sqrt{ I_1+ (1-\epsilon) I_2
}~\sqrt{-g }  d^4x.
\eeqa
In the limit $\hbar \rightarrow 0 $ the action reduces to the classical one\footnote{The general form of the perimeter action is presented in the Appendix (\ref{generalform}).}. The additional perimeter term has high derivatives of the space-time metric, and the equations of motion are much more complicated than in standard gravity. We were unable to find exact solutions of these equations and we are suggesting that  the equations can be solved by using a perturbation theory.  The classical solutions of general relativity will be modified in the regions of the space-time where the gravitational field is changing at the short-scale distances. 

Here we consider the perturbation of the Schwarzschild solution that is induced  by the the additional term in the action and try to understand how it influences the black-hole physics and the singularities. The Schwarzschild solution has the form 
\be\label{schwarz}
ds^2 = (1-{r_g \over r}) c^2 dt^2 - (1-{r_g \over r})^{-1} dr^2 -r^2 d\Omega^2~,
\ee
where 
$
g_{00} =1-{r_g \over r},~g_{11} =-(1-{r_g \over r})^{-1},~g_{22} =-r^2,~g_{33} =-r^2 \sin^2\theta,
$
and 
$$
r_g = {2GM \over c^2},~~~~ \sqrt{-g} = r^2 \sin\theta.
$$
The quadratic invariant in this case has the form
$
I_0={1 \over 12} R_{\mu\nu\lambda\rho}  R^{\mu\nu\lambda\rho} = ({r_g \over r^3})^2  
$
and shows the location of the curvature singularity at $r=0$.
 The event horizon is located where the metric component $g_{rr}$ diverges, that is,  at 
$
r_{horizon}=  r_g . 
$
The expressions for the two curvature polynomials (\ref{lineargravity2}) of our interest are\footnote{It should be stressed that on a Schwarzschild solution all other invariant polynomials presented in the Appendix of the same dimensionality  can be expressed in terms of $I_1$ and $I_2$ (\ref{massshell}).}
\beqa
I_1={9 \over 4 \pi} {r^2_g( r -  r_g )  \over  r^9},~~~~~I_2={9  \over 4 \pi}  {r^3_g \over  r^9}~, 
\eeqa
and on the Schwarzschild background the action acquires additional term of the form 
\beqa\label{lineargravityS}
S_P =   \gamma  \sqrt{{ \hbar  c^5 \over   G  }}  \int  {3    \over 2}  \sqrt{  1-\epsilon {  r_g   \over  r }}~ {r_g \over r} {dr \over r} dt~.
\eeqa
 As one can see, the expression under the square root in (\ref{lineargravityS}) becomes negative at
\be 
r < \epsilon r_g, ~~~~~0 \leq \epsilon < 1
\ee
and defines  the region where the action becomes complex and seems unreachable by the test particles.  The size of the region  depends on the parameter $\epsilon$ and is smaller than the gravitational radius $r_g$. This observation seems to have profound consequences on the gravitational singularity at $r=0$. 
In a standard interpretation of the singularities, which appear in spherically symmetric gravitational collapse,  the singularity at $r=0$ is hidden behind the event horizon. In that interpretation the singularities are still present in the theory. In the suggested 
scenario it seems possible to eliminate  the singularities from the theory based on the fundamental principles of the quantum mechanics.  The quantum-mechanical amplitude is proportional to exponent  $
 K  \sim e^{{i  \over  \hbar } S[g]} $, and  
for the Schwarzschild space-time one can find the following expression for the action:
\be\label{fullaction}
S_P =  \gamma \sqrt{{ \hbar  c^5 \over   G  }} \int^{\infty}_{r}  {3     \over 2}  \sqrt{  1-\varepsilon {  r_g   \over  r }}~ {r_g \over r} {dr \over r} \triangle t \\
= \gamma \sqrt{{ \hbar  c^5 \over   G  }}   {  1 \over \varepsilon}  \Big(  1-\big(1-\varepsilon {  r_g   \over  r }\big)^{{3\over 2}} \Big) \triangle t  . 
\ee
The action is proportional to the length $\triangle t$  of the world trajectory, as it should be for a relativistic particle at rest. In the neighbourhood of the Schwarzschild space-like singularity the action develops a large imaginary value that exponentially suppresses the propagation amplitude for the particles to enter the singularity region (\ref{amplitude}).

It is important therefore to find the perturbation of the Schwarzschild solution that is induced by the presence of the additional perimeter term in the action. The full equation has the following form: 
\beqa\label{fieldeq}
 &&{\delta S \over  \delta g^{\mu\nu}}={\delta S_A  \over  \delta g^{\mu\nu}} +  \gamma ~{\delta S_P  \over  \delta g^{\mu\nu}}  = - {  c^3 \over 16 \pi G  }   (R_{\mu\nu} - {1  \over 2} R g_{\mu\nu}) +  \\
 &&+  \gamma \sqrt{\hbar} \ \sqrt{ c^3  \over  16 \pi G}    \Big(  {1  \over 2}   {1 \over \sqrt{ I_1+ (1-\epsilon) I_2}   } ( {\delta I_1  \over  \delta g^{\mu\nu}} +  (1-\epsilon){\delta I_2  \over  \delta g^{\mu\nu}}) - {1  \over 2} \sqrt{ I_1+ (1-\epsilon) I_2}     ~g_{\mu\nu} \Big) ~=  \nn \\
 &&  - {  c^3 \over 16 \pi G  }   (R_{\mu\nu} - {1  \over 2} R g_{\mu\nu}) +  \gamma \sqrt{\hbar} \ \sqrt{ c^3  \over  16 \pi G}  \Lambda_{\mu\nu} =0,
\eeqa
where $ \Lambda_{\mu\nu} $ is a new "energy-momentum" like term induced by the perimeter perturbation.
We will search the solution of these equations in the following standard spherically symmetric form \cite{Oppenheimer:1939ue}: 
\be\label{sphericallysymm}
ds^2 = e^{\nu(r)} c^2 dt^2 -e^{\lambda(r)}dr^2 - r^2 (d\theta^2 + \sin^2\theta d\phi^2).
\ee
At the zero order in $\gamma$ the equations are:  
\beqa
R^0_0 -  {1  \over 2} R g_{0}^{0} = {   e^{-\lambda (r)} (  r \lambda^{'}(r) + e^{\lambda (r)} - 1  )    \over r^2 } =0,~~\nn\\
R^1_1 -  {1  \over 2} R g_{1}^{1} ={   e^{-\lambda (r)} ( - r \nu^{'}(r) + e^{\lambda (r)} - 1  )    \over r^2 } =0,
\eeqa
and the solution representing the Schwarzschild metric (\ref{schwarz}) is:
\be\label{BHsol}
 \nu_0 (r)  =  \log( 1- {r_g \over r}  ),~~~ \lambda_0 (r)  = - \log( 1- {r_g \over r}  ) .~~
\ee 
In order to solve the equations in the first order in $\gamma$ we  will represent the matrix  $g_{\mu\nu}$ in the form:
\be
g^{\mu\nu} = g_{0}^{ \mu\nu} + \gamma g_{1}^{ \mu\nu},~~~\nu \rightarrow  \nu_0 + \gamma \nu_1,~~~\lambda = \lambda_0 + \gamma \lambda_1,
\ee
where $g_{0}^{ \mu\nu}$ is the Schwarzschild solution. The expansion of the equations has the form: 
\beqa
&&{\delta S_A [g_0 +\gamma g_1] \over  \delta g^{\mu\nu}} + \gamma  {\delta S_P[g_0 +\gamma g_1]  \over  \delta g^{\mu\nu}} ={\delta S_A [g_0 ] \over  \delta g^{\mu\nu}}+
\gamma ~{\delta^2 S_A [g_0 ] \over  \delta g^{\mu\nu} \delta g^{\lambda\rho}}  ~   g_{1}^{\lambda\rho}+  \gamma~{\delta S_P[g_0]  \over  \delta g^{\mu\nu}} + \CO(\gamma^2) +...=0, \nn\\
\eeqa
and in the first order the  equation  is: 
\beqa\label{varequ}
&& ~{\delta^2 S_A [g_0 ] \over  \delta g^{\mu\nu} \delta g^{\lambda\rho}}  ~ g_{1}^{\lambda\rho}+   ~{\delta S_P[g_0]  \over  \delta g^{\mu\nu}}  =0.
\eeqa
The equation is linear in $g_{1}^{\lambda\rho}$ and requires the calculation of the second and first variational derivatives of the actions  $S_A$ and  $S_P$.  The second variation of the $S_A$ on the  Schwarzschild solution (\ref{BHsol}) gives :
\beqa
&& - {  c^3 \over 16 \pi G  }  {   e^{-\lambda_0 } (  (1- r \lambda^{'}_0 )\lambda_1 +r \lambda^{'}_1   )    \over r^2 } = - {  c^3 \over 16 \pi G  }[{ 1   \over r } (1-{r_g \over r} )\lambda^{'}_1 + { 1   \over r^2 } \lambda_1] ,~~\nn\\
&&- {  c^3 \over 16 \pi G  } {   e^{-\lambda_0} ( (1+r \nu^{'}_0)\lambda_1 - r \nu^{'}_1 )    \over r^2 } = - {  c^3 \over 16 \pi G  }[
 -{ 1   \over r } (1-{r_g \over r} )\nu^{'}_1 + { 1   \over r^2 } \lambda_1] .
\eeqa
The first variation of the perimeter  action $S_P$  on the Schwarzschild solution takes the form: 
\beqa\label{variation}
&& \gamma\sqrt{\frac{\hbar c^3 }{16 \pi  G}}  \left( a_1{1 \over r^3 } -a_2 {r_g \over r^4 }  + a_3 {r^2_g \over r^5 } \right)   {1\over \sqrt{1- \varepsilon  {r_g \over r} }} , \nn\\
&& \gamma \sqrt{\frac{\hbar c^3 }{16 \pi  G}}  \left( b_1{1 \over r^3 } -b_2 {r_g \over r^4 }  + b_3 {r^2_g \over r^5 } \right)   {1\over \sqrt{1- \varepsilon  {r_g \over r} }} , 
\eeqa
where the coefficients $a_1,a_2,a_3, b_1,b_3,b_3$ are given in the Appendix.
By inserting these expressions into the equation (\ref{varequ}) we will get:
\beqa
&&
{ 1   \over r } (1-{r_g \over r} )\lambda^{'}_1 + { 1   \over r^2 } \lambda_1  
= \gamma \sqrt{\hbar \frac{16 \pi  G }{  c^3 }}  \left( a_1{1 \over r^3 } -a_2 {r_g \over r^4 }  + a_3 {r^2_g \over r^5 } \right)   {1\over \sqrt{1- \varepsilon  {r_g \over r} }}, \nn\\
&&
 -{ 1   \over r } (1-{r_g \over r} )\nu^{'}_1 + { 1   \over r^2 } \lambda_1  =\gamma \sqrt{\hbar \frac{16 \pi  G}{  c^3 }}  \left( b_1{1 \over r^3 } -b_2 {r_g \over r^4 }  + b_3 {r^2_g \over r^5 } \right)   {1\over \sqrt{1- \varepsilon  {r_g \over r} }} ,
\eeqa
where 
\be
l_P= \sqrt{\hbar \frac{16 \pi  G }{  c^3 }} 
\ee
is the Planck length.  Multiplying the equations by $r^2$ yields:  
\beqa\label{mainequation}
&&
 (r-r_g   )\lambda^{'}_1 + \lambda_1  =   \gamma l_P \left( a_1{1 \over r } -a_2 {r_g \over r^2 }  + a_3 {r^2_g \over r^3 } \right)   {1\over \sqrt{1- \varepsilon  {r_g \over r} }}, \nn\\
&&
 - (r- r_g   )\nu^{'}_1 +  \lambda_1  =  \gamma l_P \left( b_1{1 \over r } -b_2 {r_g \over r^2 }  + b_3 {r^2_g \over r^3 } \right)   {1\over \sqrt{1- \varepsilon  {r_g \over r} }} .
\eeqa
The first equation can be integrated:
\beqa
&&
 (r-r_g   )\lambda_1=     \gamma  l_P \int dr  \left( a_1{1 \over r } -a_2 {r_g \over r^2 }  + a_3 {r^2_g \over r^3 } \right)   {1\over \sqrt{1- \varepsilon  {r_g \over r} }}
\eeqa
and gives: 
\beqa\label{lambda1}
\lambda_1 &=     \gamma {   l_P   \over  r-r_g   } \Big[    ~ a_1 \log{ {1 + \sqrt{1- \varepsilon  {r_g \over r} }  \over 1 - \sqrt{1- \varepsilon  {r_g \over r} } } }     
+   2{ a_3 - a_2 \varepsilon \over \varepsilon^2 } \sqrt{1- \varepsilon  {r_g \over r} }   
-  {2 a_3  \over 3 \varepsilon^2 }  (1 - \varepsilon  {r_g \over r} )^{3/2}  + Const.  \Big]\nn\\
&\equiv   \gamma {  l_P   \over  r-r_g   }  \Big[ f\Big({r_g \over r}\Big) +Const.\Big]\nn\\
\eeqa
At $r \rightarrow \infty$ $\lambda_1$ has the following asymptotic: 
\be
\lambda_1~ \simeq ~ \gamma {  l_P   \over  r  } 
\Big[   a_1   \log{ 4 r  \over   \varepsilon  r_g  }     
-    { 2 a_2 \over \varepsilon}   +    {4 a_3 \over 3 \varepsilon^2}     \Big] +\CO({1\over r^2}), 
\ee
and at the horizon $r \rightarrow r_g$ it has the form: 
\be
\lambda_1 ~\simeq ~ \gamma { l_P   \over  r-r_g   }  
\Big[   a_1  \log{ {1 + \sqrt{1- \varepsilon  }  \over 1 - \sqrt{1- \varepsilon  } } }       
-   2 { a_3 - a_2 \varepsilon \over \varepsilon^2 }   \sqrt{1- \varepsilon  }   
+ {2 a_3  \over 3 \varepsilon^2} (1- \varepsilon )^{3/2}     +\CO(r-r_g) \Big].
\ee
In order to get a standard  behaviour of the solution near the horizon and at infinity one should subtract the last term appropriately choosing   the integration constant in (\ref{lambda1})\footnote{I would like to thank Konstantin for suggesting the above boundary condition.}. The $g_{11}$ component of the metric now becomes equal to the following expression: 
\beqa\label{lambda11}
&&g_{11}=e^{\lambda_0 + \lambda_1}= {-1 \over 1-  {r_g \over r}   }   ~e^{  \lambda_1} ,
\eeqa
where 
\beqa\label{lambda111}
  \lambda_1 &=&        \gamma  { l_P   \over  r-r_g   } \Big[  f\Big({r_g \over r}\Big)- f\Big(1\Big)\Big].
\eeqa
With this choice of the integration constant we will get the following leading behaviour of the metric $g_{11}$ at infinity:
\beqa\label{g11infty}
g_{11} ~\simeq ~ {-1 \over 1-  {r_g \over r}   }  
&& \exp{     \Big[   \gamma~ {l_P  \over  r  }  \Big(       a_1   \log{  r  \over   r_g  } +a \Big)~ \Big]}  ,
~~~~~
\eeqa
where the coefficients are $a_1 = {82  \over 5 \sqrt{\pi}} + \CO( \varepsilon ),  a=-{33 \over \sqrt{\pi} }  + \CO( \varepsilon )$.  The behaviour near the horizon is:
\beqa\label{g11horizon}
&&g_{11} ~\simeq ~ {-1 \over 1-  {r_g \over r}   }  
 \exp{     \Big[   \gamma~ {l_P  \over  r_g  }   b \Big]   }, 
 \eeqa
where $b = (a_1-a_2+ a_3)/ \sqrt{1- \varepsilon  } =  {11 \over 10 \sqrt{\pi} }  + \CO( \varepsilon )$. In order to find the time component of the metric $g_{00}$ we subtract the second equation from the first one in (\ref{mainequation}):
\beqa\label{mainequationnu}
&&
 (r-r_g   )(\lambda^{'}_1 +\nu^{'}_1)   =   \gamma l_P \left( c_1{1 \over r } -c_2{r_g \over r^2 }  + c_3 {r^2_g \over r^3 } \right)   {1\over \sqrt{1- \varepsilon  {r_g \over r} }},
\eeqa
where the coefficients $c_1,c_2,c_3$ are given in Appendix.
Thus
\beqa\label{nu1eq}
 \nu_1  + \lambda_1  =  \gamma l_P \int \left( c_1{1 \over r } -c_2{r_g \over r^2 }  + c_3 {r^2_g \over r^3 } \right)    {d r \over (r-r_g   ) \sqrt{1- \varepsilon  {r_g \over r} }} ,
\eeqa
and the integration gives  
\beqa\label{thenu1}
\nu_1 + \lambda_1 =         
   \gamma {  l_P   \over r_g   }
 \Big[   2{ c_1 \varepsilon - c_3  \over   \varepsilon^2  } \sqrt{1- \varepsilon  {r_g \over r} }   +  {2 c_3 \over 3 \varepsilon^2 } (1  - \varepsilon  {r_g \over r} )^{3/2} + Const \Big ] \equiv   \gamma{  l_P   \over r_g   } \Big[g\Big({r_g \over r}\Big) + Const  \Big], ~~~~~~~   
\eeqa
where we used the important relation $c_1 - c_2 +c_3 =0$ that  eliminates a logarithmic singularity in $\nu_1$  (see Appendix). The $\lambda_1$ is given in  (\ref{lambda111}).
We should choose the integration constant equal to the asymptotic value of the integral at the infinity:
\beqa
&& \nu_1 +\lambda_1 ~\simeq ~ \gamma {  l_P   \over r_g   }
 \Big[    2{ c_1 \varepsilon - c_3  \over   \varepsilon^2   }  +  {2 c_3 \over 3 \varepsilon^2 } \Big]. \nn
\eeqa
Thus the time component of the metric is equal to the following expression: 
\beqa\label{gamma00}
&&g_{00}=e^{\nu_0 + \nu_1}  =   (1- {r_g \over r} )   e^{  \nu_1} ~,\nn\\
&&\\
\nu_1=    
 -  \gamma {  l_P   \over  r-r_g   }  &&\Big[  f \Big({r_g \over r} \Big)- f(1)\Big]+   \gamma {   l_P   \over r_g   } \Big[g\Big({r_g \over r} \Big)-g\Big(0\Big)\Big].  \nn    
\eeqa
With this choice of the integration constant we will get the following leading behaviour of the metric $g_{00}$ at infinity: 
\beqa\label{g00infty}
g_{00} ~\simeq ~ (1- {r_g \over r} ) 
 \exp{    \Big[  - \gamma~ {l_P  \over  r  }   \Big(   a_1   \log{  r  \over   r_g  } + a +c_1 \Big)~ \Big]} ,
 \eeqa 
where $c_1 = {41 \over 3 \sqrt{\pi}} + \CO( \varepsilon)$  and near the horizon as
\beqa\label{g00horizon}
g_{00} ~\simeq ~ && (1- {r_g \over r} )   \exp{     \Big[  - \gamma~ {l_P  \over  r_g  }  (   b +d )  \Big] },      
\eeqa
where $d=-{2 \over 3 \sqrt{\pi}}  +\CO(\varepsilon)$. 

In summary, the spherically symmetric  metric (\ref{sphericallysymm}) is given by the formulas  (\ref{lambda11})  and (\ref{gamma00}).  The asymptotics of the metric components at infinity are given by (\ref{g11infty})  and (\ref{g00infty}), and the asymptotics of the metric components near the horizon are given in (\ref{g11horizon}) and (\ref{g00horizon}). From the obtained solution it is clearly seen that the characteristic behaviour of the 
corrections to the Schwarzschild solution is defined by the irrational functions of 
the form (\ref{lambda1}) and (\ref{thenu1})
\be
 ~\sum^{\infty}_{n=0} \alpha_n (1- \varepsilon  {r_g \over r}  )^{n/2} 
\ee 
that are developing the imaginary parts when $r < \varepsilon  r_g $ and generate a something like a firewall \cite{Almheiri:2012rt}  prohibiting particles entering the singularity. Let us compare this behaviour with the behaviour of the amplitude (\ref{path}) for a relativistic particle 
$ \triangle K =   e^{ {i \over \hbar}  m c^2 \sqrt{1- {v^2 \over c^2}}\triangle t} $.
For the velocities larger than the velocity of light the amplitude is exponentially decreasing and the propagation of a particle outside of the light-cone is suppressed as one can see in Fig.\ref{fig1}.  The evaluation of the path integral leads to the Feynman propagator for relativistic scalar particle  \cite{Feynman:1949hz,Feynman:1949zx}.
\begin{figure}
\begin{center}
\includegraphics[width=8cm]{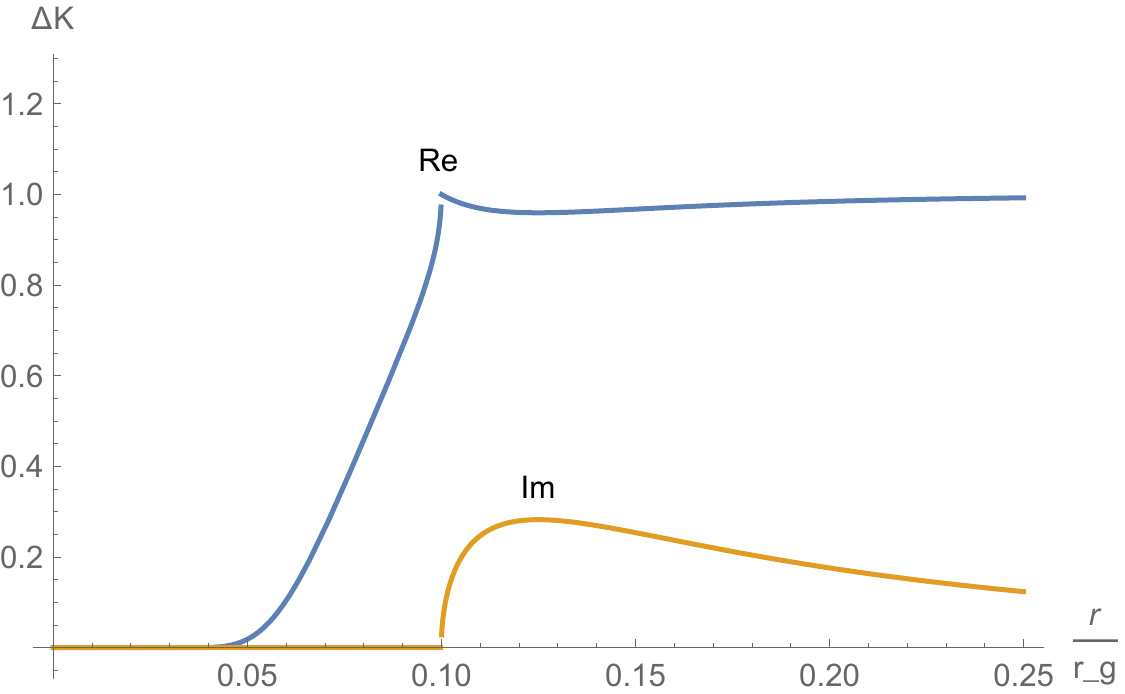}~~~~~~~~~~
\caption{
The graph of the real and imaginary parts of the   amplitude  in gravity with  perimeter action (\ref{actionlineargravity23})  $ \triangle K = e^{ i  \gamma  \sqrt{{  c^5 \over    \hbar G  }}   {3   \over 2}      \sqrt{  1-\epsilon {  r_g   \over  r }}~ {r_g \over r} {\triangle r \over r} \triangle t}$, here $\epsilon =0.1$ and $\gamma  \sqrt{{  c^5 \over    \hbar G  }} {3   \over 2} { \triangle r  \over r_g} \triangle t =0.01$.   
}
\label{fig2}
\end{center}
\end{figure} 
In the case of gravity with perimeter action the same phenomena appears in the region $r < \varepsilon r_g$ where the  action (\ref{actionlineargravity23}), (\ref{lineargravityS}) is developing an imaginary part shown in  Fig.\ref{fig2} and the expectation value of any observable $\langle  \CO \rangle$ in the region $r < \varepsilon r_g$ is exponentially suppressed.    

Let us also consider the behaviour of the solution at infinity and near the horizon.
As we saw, the perturbation (\ref{actionlineargravity23}) generates a deformation to the distance invariant $ds$  (\ref{schwarz}) and allows to calculate  the correction to the temporal component of the metric tensor  caused by the additional term in the  action at infinity and near the horizon.  From (\ref{g00infty}) we will get 
the behaviour of the temporal component of the metric at large distances $r \gg r_g$:
\beqa\label{main}
&&g_{00} = 1  + 2 {\phi \over c^2} ~\simeq ~ 1-{r_g \over r} 
 - \gamma~ {l_P  \over  r  }     \Big(  a_1   \log{  r  \over    r_g  } +a+ c_1 \Big),
\eeqa
with the additional logarithmic correction to the gravitational potential.  The parameters  appearing in this expression  are: 
$$
l_P= \sqrt{\hbar \frac{16 \pi  G }{  c^3 }} ~\simeq ~
11.3 \times 10^{-33} cm ,~~~a_1 = \frac{82 }  {5 \sqrt{\pi }} +\CO( \varepsilon), ~a =  -{33 \over \sqrt{\pi }} +\CO( \varepsilon),~c_1 = \frac{41 }  { 3 \sqrt{\pi } }+\CO( \varepsilon ).\nn
$$
The additional attractive term in the potential $\phi$ is logarithmically  increasing with distance. 
The correction  $ \gamma~ {l_P  \over  r  }       \log{  r  \over    r_g  } $ is tiny  
because  the action (\ref{lineargravity1}) contains the Planck constant  $\hbar$ in front of the action and the mass parameter is proportional to the Planck mass (\ref{Planckmass}).  For the most astrophysical bodies  the ratio $l_P/r_g \ll 1$  is very small\footnote{One can consider a larger mass parameter $M$ in (\ref{lineargravity}) in order to accommodate a flat rotation curve of spiral galaxies. The logarithmically growing potential in (\ref{main}) will increase the rotation velocities. }. 
The potential of Sun on the Earth orbit will receive the following correction: 
$
g_{00}(M\textsubscript{\(\odot\)}) ~\simeq 1 - 10^{-8} - \gamma ~10^{-43}
$,
and the potential generated by Milky Way on the Sun orbit will be of the order 
$
g_{00}(M\textsubscript{\(MW\)}) ~\simeq 1 - 10^{-5} - \gamma~ 10^{-54},
$
where  $\gamma$ is the coupling constant in (\ref{actionlineargravity23}).   The  gravitational time dilation near a massive body also receives a tiny correction, and therefore $ d \tau   \leq dt $ as in the standard gravity. The advance precession of the 
perihelion $\delta \phi$  expressed in radians per revolution is  
\be
\delta \phi =  {3 \pi m^2 c^2 r^2_g  \over 2 L^2 } \Big(1 + 2 \gamma {l_P  \over r_g} (8 a_1+a+ c_1 )\Big)=
{6 \pi G M   \over c^2 l (1-e^2) }\Big(1 + 2 \gamma {l_P  \over r_g} (8 a_1+a+ c_1 )\Big),
\ee
where $l$ is the semi-major axis and $e$ is the orbital eccentricity.  The precession is advanced by the additional factor $  \gamma {l_P  \over r_g}   $   is tiny compared with the 
experimental uncertainty  in the observational data for the advanced precession of the Mercury perihelion,  which is $42,98 \pm 0,04$ seconds of arc per century. From the equation for light-like geodesics and from (\ref{g00horizon}), (\ref{g11horizon}) it follows that the horizon remains undisturbed and is at $r=r_g$.

In conclusion I  would like to thank Jan Ambjorn for the invitation  and
kind hospitality in the Niels Bohr Institute, where part of the work was done. 
I  would like to thank Alex Kehagias,  Kyriakos Papadodimas and Konstantin Savvidy for stimulating 
discussions. The author acknowledges support by the ERC-Advance Grant 291092,  "Exploring the Quantum Universe" (EQU).  

\section{\it Appendix}

The general expression for the perimeter action has the form: 
\beqa\label{generalform}
 && S_P = 
   \sqrt{\hbar} \ \sqrt{ c^3  \over  16 \pi G}  \int  \sqrt{\sum^{3}_{1}  \eta_i K_i+\sum^{4}_{1}  \chi_i J_i+  \sum^{9}_{1}  \gamma_i I_i}~\sqrt{-g }  d^4x ~,
\eeqa
where the curvature invariants have the form 
\beqa
&&I_0={1\over 12} R_{\mu\nu\lambda\rho}  R^{\mu\nu\lambda\rho}, ~~
 I_1= -{1\over 80 \pi }R_{\mu\nu\lambda\rho;\sigma} R^{\mu\nu\lambda\rho;\sigma},~~
I_2= {1\over 16 \pi } R_{\mu\nu\lambda\rho}  \Box  R^{\mu\nu\lambda\rho}~,\nn\\
&&I_3= -{1\over 32 \pi} \Box (R_{\mu\nu\lambda\rho} R^{\mu\nu\lambda\rho} ),~
I_4= -{1\over 40 \pi} R_{\mu\nu\lambda\rho;\alpha} R^{\alpha\nu\lambda\rho;\mu},~
I_5=  - {1\over 8 \pi}  (R^{\alpha\nu\lambda\rho}  R^{\mu}_{~~\nu\lambda\rho})_{;\mu;\alpha}, \nn\\
&&I_6=  -{1\over 8\pi }  (R^{\alpha\nu\lambda\rho}  R^{\mu}_{~~\nu\lambda\rho})_{;\alpha;\mu},~
I_7=  {1\over 8 \pi}  R^{\alpha\nu\lambda\rho} R^{\mu}_{~\nu\lambda\rho;\alpha;\mu},~
I_8= {1\over 8 \pi}  R^{\mu}_{~\nu\lambda\rho;\mu} R^{\sigma\nu\lambda\rho}_{~~~~~;\sigma}~,  \nn\\
&&I_9=  {1\over 8 \pi}  R^{\alpha\nu\lambda\rho} R^{\mu}_{~\nu\lambda\rho;\mu;\alpha} ~,~~\nn\\
&&J_0=  R_{\mu\nu }  R^{\mu\nu } ~,~~J_1= R_{\mu \nu;\lambda} R^{\mu\nu;\lambda }~,~~J_2=  R^{\mu\nu }   \Box R_{\mu\nu }~,~~J_3=  \Box (R^{\mu\nu }    R_{\mu\nu })~,~~J_4=   R_{\mu \sigma}^{~~~;\mu} R^{\nu\sigma }_{~~~;\nu} \nn\\
&& K_0=  R^2~,~~
 K_1=  R_{;\mu}  R^{;\mu}~,~~
 K_2=  R    \Box R~,~~
 K_3=  \Box  R^2=2K_1 +2K_2~  . 
\eeqa
The $\eta_i , \chi_i $ and $ \gamma_i$ are free parameters. Some of the invariants can be expressed through others using covariant differentiation and Bianchi identities. On the Schwarzschild solution (\ref{schwarz}) all these  invariants can be expressed in terms of  the $I_1$ and $I_2$
\be\label{massshell}
I_3=I_5=I_6=5 I_1 -I_2,~~I_4=I_1,~~I_7=I_2,~~I_8=I_9=0,~~ J_i=0,~~ K_i=0.
\ee
The integrand of the perimeter action (\ref{lineargravity1}) on the Schwarzschild solution has the form:
\be
 \sqrt{ I_1+ (1-\epsilon) I_2}\vert_{ (\nu_0,\lambda_0)  } = {3   \over 2 \sqrt{\pi}  }  \sqrt{ 1  - \varepsilon {r_g\over r} }~ {  r_g \over   r^4}
\ee
and the variational derivatives of the  invariants $I_1$ and $I_2$ on the Schwarzschild solution appearing in the equation (\ref{varequ}) are:
\beqa
{\delta I_1  \over  \delta  \nu(r)}\vert_{ g^{\mu\nu}_0 } = \frac{3   r_g \left(24 r^2-77 r  r_g+54  r_g^2 \right)}{10 \pi  r^9},~
{\delta I_1  \over  \delta  \lambda(r)}\vert_{ g^{\mu\nu}_0 } =\frac{3    r_g \left(8 r^2-39 r r_g+33 r_g^2\right)}{20\pi  r^9},\nn\\
{\delta I_2  \over  \delta  \nu(r)}\vert_{ g^{\mu\nu}_0 } = \frac{3     r_g \left(56 r^2-175 r  r_g+120  r_g^2\right)}{4 \pi r^9},~
{\delta I_2  \over  \delta  \lambda(r)}\vert_{ g^{\mu\nu}_0 } =\frac{   r_g \left(28 r^2-118 r  r_g+93  r_g^2\right)}{4 \pi  r^9} .\nn
\eeqa
The above expressions allow to calculate the coefficients appearing in (\ref{variation})
\beqa
&a_1 = \frac{328-280 \varepsilon }  { 20 \sqrt{\pi } }, ~a_2 =   \frac{1014 -875 \varepsilon  }{20 \sqrt{\pi }  },~a_3 =   \frac{708-615\varepsilon }{20 \sqrt{\pi } },\nn\\
&b_1 =\sqrt{\pi } \frac{82 - 70  \varepsilon }  { 30 \sqrt{\pi } }, ~b_2 =    \frac{331 - 295 \varepsilon   }{30 \sqrt{\pi }  },~b_3 =    \frac{282 - 255 \varepsilon }{30 \sqrt{\pi } }.
\eeqa 
The integration of the equation (\ref{nu1eq})  defining the $\nu_1$ function gives:
\beqa\label{singular}
 &&
\nu_1      = - \lambda_1  
 +  \gamma {  l_P   \over r_g   }
 \Big[   - 2 { (c_1 -c_2 +c_3) \over   \sqrt{1-\varepsilon}  } \log{   {   \sqrt{1-\varepsilon} +  \sqrt{1- \varepsilon {r_g \over r}}     \over  {   \sqrt{1-\varepsilon} -  \sqrt{1- \varepsilon {r_g \over r}}   }  } }
 + \nn\\
&&  +  2{c_2\varepsilon - c_3 \varepsilon -c_3 \over   \varepsilon^2  } \sqrt{1- \varepsilon  {r_g \over r} }  +  {2 c_3 \over 3 \varepsilon^2  } (1  - \varepsilon  {r_g \over r} )^{3/2}    \Big],    
\eeqa 
where the coefficients $c_1,c_2,c_3$  are:
\beqa
&&c_1 = a_1-b_1=\frac{41-35  \varepsilon }  { 3 \sqrt{\pi } },~~
c_2 = a_2-b_2= \frac{476 -407 \varepsilon  }{12  \sqrt{\pi }},~~\nn\\
&&c_3 = a_3-b_3= \frac{104-89  \varepsilon }{4 \sqrt{\pi } },~~c_1-c_2+c_3 =0.
\eeqa
 The last relation allows to eliminate the singular logarithmic term in (\ref{singular}) and equation reduces 
 to the (\ref{thenu1}). 
 The coefficient $a$  in the formula (\ref{g11infty}) was obtained in the expansion:
\beqa
&& {4 a_3 \over 3 \varepsilon^2} -    { 2 a_2 \over \varepsilon}      - a_1  \log{ {1 + \sqrt{1- \varepsilon  }  \over 1 - \sqrt{1- \varepsilon  } } }       
-    {2 a_3 -2 a_2 \varepsilon \over \varepsilon^2 }   \sqrt{1- \varepsilon  }   
+ {2 a_3  \over 3 \varepsilon^2} (1- \varepsilon )^{3/2}+a_1   \log{  4 r  \over    \varepsilon  r_g  }= \nn\\
&& =a_1   \log{  r  \over     r_g  }  + a ~ 
+ \CO( \varepsilon ),
\eeqa
where
\be 
a_1=  \frac{82 }{5   \sqrt{\pi } }+ \CO( \varepsilon ),~~~~~a= { a_3  - 2 a_2 \over 2} =-{33  \over  \sqrt{\pi }}  + \CO( \varepsilon )
\ee
The coefficient $d$ in (\ref{g00horizon}) is: 
\be
d=     2{ c_1 \varepsilon - c_3  \over   \varepsilon^2    }(\sqrt{1- \varepsilon    }-1)  + {2 c_3 \over 3 \varepsilon^2 } ( (1  - \varepsilon  )^{3/2}-1)  \simeq   { c_3 -2c_1\over 2 }   +\CO(\varepsilon) =
-{2\over 3 \sqrt{\pi }}   +\CO(\varepsilon). 
\ee 
The field equations (\ref{fieldeq}) can be represented also in the standard form:
\beqa
R_{\mu\nu}- {1\over 2} R g_{\mu\nu}= \gamma \sqrt{\hbar \frac{16 \pi  G }{  c^3 }}  \Lambda_{\mu\nu} + 
\frac{16 \pi  G }{  c^4 } T_{\mu\nu},
\eeqa
where  $ \Lambda_{\mu\nu} $ is a new "energy-momentum" like term induced by the perimeter perturbation and the $T_{\mu\nu}$ is the standard energy-momentum tensor of matter.

The path integral for a relativistic particle amplitude can conveniently be represented in the form: 
\be\label{path}
K(t_b,x_b;t_a,x_a)= \int^{x_b}_{x_a}  e^{ {i \over \hbar} m c^2 \int^{t_b}_{t_a} \sqrt{1- {\dot{x}^2 \over c^2}}dt} \CD x(t).
\ee
 
 \section{\it Note Added} 
In order to investigate the perturbation of the Schwarzschild solution induced by the perimeter action  in its general form (\ref{generalform}) one should calculate the variational derivatives of the invariants $I_i$  that appear in the equation (\ref{varequ}) on the Schwarzschild solution. The general relations between invariants $I_i$ are:
\be
I_3 = 5 I_1 -I_2,~~~I_6 =I_5 = 5I_4 - I_7 - I_8 - I_9~,~~
\ee
therefore the independent invariants are: $I_1,I_2, I_4, I_7,I_8,I_9$.  Now the field equation (\ref{fieldeq}) will take the form: 
\beqa\label{fieldeq1}
{\delta S \over  \delta g^{\mu\nu}}={\delta S_A  \over  \delta g^{\mu\nu}} &+&  \gamma ~{\delta S_P  \over  \delta g^{\mu\nu}}  = - {  c^3 \over 16 \pi G  }   (R_{\mu\nu} - {1  \over 2} R g_{\mu\nu}) +  \\
 &&+   \sqrt{\hbar} \ \sqrt{ c^3  \over  16 \pi G}    \Big(  {1  \over 2}   {1 \over \sqrt{  \gamma_i I_i}   }  ~\gamma_j  {\delta I_j  \over  \delta g^{\mu\nu}}   - {1  \over 2} \sqrt{  \gamma_i I_i}       ~g_{\mu\nu} \Big)=0 , ~~~i,j=1,2,4,7,8,9, \nn
\eeqa
where the summation is over all possible invariants $I_i$. 
The first variational derivatives of these invariants on the Schwarzschild solution are:
\beqa\label{variations}
&{\delta I_4  \over  \delta  \nu(r)}\vert_{ g^{\mu\nu}_0 } = {\delta I_1  \over  \delta  \nu(r)}\vert_{ g^{\mu\nu}_0 } ,~
&{\delta I_4  \over  \delta  \lambda(r)}\vert_{ g^{\mu\nu}_0 } ={\delta I_1  \over  \delta  \lambda(r)}\vert_{ g^{\mu\nu}_0 } ,\nn\\
&{\delta I_7  \over  \delta  \nu(r)}\vert_{ g^{\mu\nu}_0 } = {\delta I_2  \over  \delta  \nu(r)}\vert_{ g^{\mu\nu}_0 },~
&{\delta I_7  \over  \delta  \lambda(r)}\vert_{ g^{\mu\nu}_0 } ={\delta I_2  \over  \delta  \lambda(r)}\vert_{ g^{\mu\nu}_0 } ,\\
&{\delta I_8  \over  \delta  \nu(r)}\vert_{ g^{\mu\nu}_0 } = 0,~
&{\delta I_8  \over  \delta  \lambda(r)}\vert_{ g^{\mu\nu}_0 } =0 \nn,\\
&{\delta I_9  \over  \delta  \nu(r)}\vert_{ g^{\mu\nu}_0 } = \frac{3    r_g \left(112 r^2-301 r  r_g+192  r_g^2 \right)}{8 \pi r^9},~
&{\delta I_9  \over  \delta  \lambda(r)}\vert_{ g^{\mu\nu}_0 } =\frac{  r_g \left(56 r^2-143 r r_g+96 r_g^2\right)}{8 \pi r^9}.\nn
\eeqa
The invariants $J_i$ and $K_i$ do not contribute. Taking into account the relations (\ref{massshell}) and (\ref{variations}) one can represent the second term  of the perturbation equation (\ref{varequ}) on the Schwarzschild solution in the form: 
\beqa\label{vareeq1}
 &&\gamma   \sqrt{\hbar} \ \sqrt{ c^3  \over  16 \pi G}    \Big(  {1  \over 2}   {1 \over \sqrt{  I_1 + (1-\varepsilon) I_2}   }  ~  \Big(  {\delta I_1  \over  \delta g^{\mu\nu}}    + (1-\varepsilon)   {\delta I_2  \over  \delta g^{\mu\nu}}  + \eta   {\delta I_9  \over  \delta g^{\mu\nu}}  \Big)  - {1  \over 2}    \sqrt{  I_1 + (1-\varepsilon) I_2  }    ~g_{\mu\nu} \Big) ,  \nn
 \eeqa
where 
\be 
\gamma = \sqrt{\gamma_1 + \gamma_4},~~~1-\varepsilon = {\gamma_2 + \gamma_7 \over \gamma_1 + \gamma_4},
~~~\eta = { \gamma_9 \over \gamma_1 + \gamma_4},~~~\gamma_1 + \gamma_4 >0.
\ee
Thus compared to the perimeter perturbation that we considered in the main part of the article we have here an additional perturbation term with the coupling constant $\eta$. As one can see from the last relation in (\ref{variations}), this term has the same dependence on the radial coordinate $r$ as the terms associated with 
$I_1,I_2$, therefore the consideration of the general perimeter perturbation reduces to the following redefinition of the coefficients $a_i, b_i$ and $c_i$: 
\beqa\label{newcoefficients}
&a_1 = \frac{328-280 \varepsilon +280 \eta}  { 20  \sqrt{\pi }}, ~&a_2 =    \frac{1014 -875 \varepsilon  +1505 \eta/2 }{20   \sqrt{\pi }},~a_3 =    \frac{708-615\varepsilon +480 \eta}{20  \sqrt{\pi }},\nn\\
&b_1 =  \frac{82 - 70  \varepsilon +70  \eta}  { 30 \sqrt{\pi } }, ~&b_2 =    \frac{331 - 295 \varepsilon  +715 \eta/4 }{30  \sqrt{\pi } },~b_3 =   \frac{282 - 255 \varepsilon +120 \eta}{30  \sqrt{\pi }}.
\eeqa 
\begin{figure}
\begin{center}
\includegraphics[width=8cm]{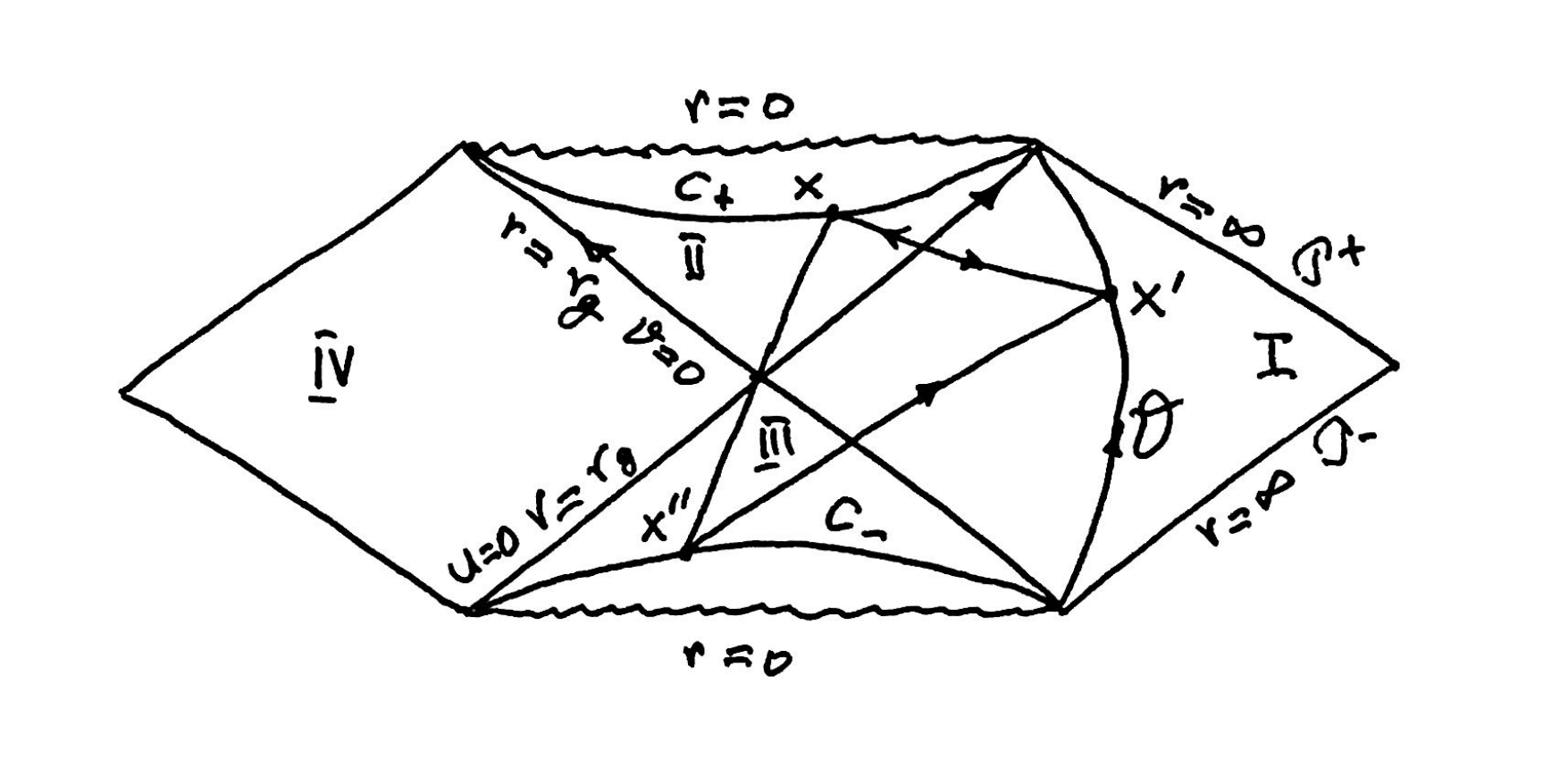}~~~~~~~~~~
\caption{
The Penrose-Carter diagram of the Schwarzschild space-time  \cite{Hartle:1976tp}. The horizontal lines $r=0$ are the future and past singularities. The diagonal lines bounding the diagram on the right-hand side are past and future null infinity of asymptotically flat space.  The diagonals are the past and future horizons. The  propagator (\ref{emission})  represents the propagation from point $x$ on the surface $C_+$ of $r = constant < r_g$ in the future horizon $II$ to the point $x^{'}$ on the world line of the observer $\CO$ on the surface  $r^{'}=constant > r_g$. The integration over  time coordinate can be shifted by the amount $-2\pi i r_g$ in the complex $t$ plane (\ref{reflection}) and corresponds to the reflection of the point $x$ in the origin to the point $x^{''}$ in the region $III$ on the surface $C_{-}$ of  $r^{''} = constant < r_g$.  The resulting amplitude describes the propagation to the  point  $x^{'} = (0,\vec{r}{~'})$ outside the black hole from the surface $C_-$ in the past horizon region $III$. The time reversal transformation of the amplitude (\ref{reflection}) represents the propagation of massless particle from the region $I $ into the interior region $II$ inside the black hole. 
}
\label{fig3}
\end{center}
\end{figure} 
Considering the massless scalar particles in the perturbed Schwarzschild background we can ask: What is the amplitude of finding the particles at location $x$ inside the future horizon? To describe this situation we shall follow the Hartle and Hawking path integral derivation of the amplitude for a particle that propagates in the Schwarzschild space-time background \cite{Hawking:1974sw,Hartle:1976tp,Gibbons:1976ue,Gibbons:1978ac}.  Let us consider the amplitude for a massless scalar particle of defined energy $E$ to propagate from the space-time point $x^{\mu} =(t,\vec{r} )$  inside the region $II$ of the future horizon to the point $x^{' \mu} =(t^{'},\vec{r}{~'} )$ outside the black hole in the region $I$ shown in Fig.\ref{fig3}:  
\be\label{emission}
K_{E}(\vec{r}{~'}, \vec{r} )= \int_{-\infty}^{+\infty} d t e^{-i E t} K(0,\vec{r}{~'}; t,\vec{r} ),
\ee
where the integration is over time $t$ inside the future horizon and $\vec{r}$ is on a surface $C_+$ of $r = constant < r_g$ \cite{Hartle:1976tp}. 
Because the propagator is symmetric with respect to the coordinate exchange  $K(x^{'},x)= K(x,x^{'})$ the amplitude can be represented in the form  \cite{Hartle:1976tp}: 
\be
K_{E}( \vec{r}, \vec{r}{~'} ) = \int_{-\infty}^{+\infty} d t e^{-i E t} K( t,\vec{r};0,\vec{r}{~'} ).
\ee
The propagator $K( t,\vec{r};0,\vec{r}{~'} )$ is analytic in the coordinate $t$ in the strips of the width $2\pi r_g$ except  for the singularities that are below the real $t$ axis and  correspond to propagation along the future-directed null geodesics,   while those corresponding to propagation along the past-directed null geodesics lie above the real $t$ axis  \cite{Hartle:1976tp}.  By distorting the contour of the $t$ integration downward by amount $-2\pi i r_g$ in the complex $t$ plane, one can represent the amplitude in the form \cite{Hartle:1976tp}:
\be\label{reflection}
K_{E} (\vec{r}{~'}, \vec{r} )= \int_{-\infty}^{+\infty} d t e^{-i E (t-2\pi i r_g)} K( t-2\pi i r_g,\vec{r};0,\vec{r}{~'} ).
\ee
Since the displacement in $t= \tau + i \sigma$ by $\sigma \rightarrow  \sigma -2\pi i r_g$  is equivalent to the reflection $u \rightarrow -u$ and $v \rightarrow -v$ of the Kruskal coordinates (  $u= \vert u \vert e^{-i {\sigma \over 2 r_g}}$ and $v= \vert v \vert e^{i {\sigma \over 2 r_g}}$ )  the last integral can be interpreted as the amplitude to propagate to the point $x^{'} = (0,\vec{r}{~'})$  outside the  black hole, but now from the surface $C_-$  in the past horizon $III$  that is the reflection of the $C_+$ surface in the origin of the Kruskal $(u,v)$ coordinates defined above. Therefore the last amplitude may be written  as \cite{Hartle:1976tp}:
\be\label{absoption}
K_{E}(\vec{r},\vec{r}{~'} ) =e^{-2\pi E r_g} \int_{-\infty}^{+\infty} d t e^{-i E t} K( t,\vec{r};0,\vec{r}{~'} ),
\ee
where now the integration is over time $t$ inside the past horizon $III$ and the $\vec{r}$ is on the reflected surface $C_-$ of $r = constant < r_g$ \cite{Hartle:1976tp}. It represents the amplitude to propagate to the point  $x^{'} = (0,\vec{r}{~'})$ outside the black hole from the surface $C_-$ in the past horizon region $III$ of the Penrose-Carter diagram Fig.\ref{fig3}. By time-reversal invariance this amplitude is exactly equal to the amplitude for a particle that starts at $x^{'} = (0,\vec{r}{~'})$ and arrives at $C_+$ inside the black hole. This is exactly the amplitude of finding the particles at the location $x$ inside the future horizon.

Now we are able to calculate the amplitude of a particle to reach the region $r < \varepsilon r_g$ inside the future horizon under the perimeter perturbation of the Schwarzschild space-time.  This amplitude is represented by the path integral 
\beqa\label{amplitude}
 A_E(r < \varepsilon r_g)&=&\int^{r< \varepsilon r_g}_{0} d^3 \vec{r} \int^{x}_{x^{'}} e^{ {i\over \hbar} (S[g] + S[g,x])}  \CD g  \CD x   ~\simeq ~  \int^{r< \varepsilon r_g}_{0} d^3 \vec{r} \int^{x}_{x^{'}} e^{{i\over \hbar} S[g_0+\gamma g_1] + S[g_0,x]}  \CD x = \nn\\
 =  &&  \int^{r< \varepsilon r_g}_{0} d^3 \vec{r} ~e^{{i\over \hbar} S_P} K_{E}(\vec{r},\vec{r}{~'} )=   \int^{r< \varepsilon r_g}_{0} d^3 \vec{r} ~e^{- \gamma     \sqrt{ {  \varepsilon  c^5 \over  \hbar G}  }   \big(  {r_g\over r} \big)^{3/2} \triangle t } K_{E}(\vec{r},\vec{r}{~'} )
\eeqa
and due to the exponential factor $\exp{\big(-C \big(  {r_g\over r} \big)^{3/2} \big)}$  the  amplitude tends to zero when a massless scalar particle  approaches the singularity $r=0$.   The  $S_P$ is given in (\ref{lineargravityS}) and (\ref{fullaction}).  The Feynman propagators are known for the photons and fermions \cite{Hartle:1976tp}  and for the different choices of boundary conditions \cite{Candelas:1980zt}. These results allow to calculate the corresponding amplitudes and to conclude that the same phenomenon will take place in the case of other elementary particles as well.

\vfill
\end{document}